\documentstyle[12pt,psfig]{article}
\newlength{\abstractwidth}
\begin{document}

\thispagestyle{empty}
\pagestyle{plain}
%%%%%%%%%%%%%%%%%%%%%%%%%%%%%%%%%%%%%%%%%%%%5
%macros%%%%%%%
%%%%%%%%%%%%%%%%%%%%%%%%%%%%%%%%%%%%%%%%%%%%5
\renewcommand{\thefootnote}{\fnsymbol{footnote}}
\renewcommand{\thanks}[1]{\footnote{#1}} % Use this for footnotes
\newcommand{\starttext}{
\setcounter{footnote}{0}
\renewcommand{\thefootnote}{\arabic{footnote}}}
\newcommand{\be}{\begin{equation}}
\newcommand{\ee}{\end{equation}}
\newcommand{\PSbox}[3]{\mbox{\rule{0in}{#3}\includegraphics{#1}\hspace{#2}}}
%%%%%%%%%%%%%%%%%%%%%%%%%%%%%%%%%%%%%%%%%%%%%%%%%%%%%%%%%%%%%%%

\begin{titlepage}
\bigskip
\hskip 3.7in\vbox{\baselineskip12pt
\hbox{MIT-CTP-2688}\hbox{hep-th/9711051}}
\bigskip\bigskip\bigskip\bigskip

\centerline{\large \bf A note on the interactions of compactified D-branes.}

\bigskip\bigskip
\bigskip\bigskip

\centerline{\bf Alec Matusis\thanks{alec\_m@ctp.mit.edu}}
\medskip
\centerline{Center for Theoretical Physics}
\centerline{Massachusetts Institute of Technology}
\centerline{Cambridge, MA\ \ 02139}

\bigskip\bigskip

\begin{abstract}
\baselineskip=16pt
We compute the vacuum one loop amplitude for two D-strings at an angle compactified on $T^2$ as a function of the transverse separation and the winding numbers.
We show that in certain cases the  amplitude is independent on whether the D-strings are compactified or not. 
\end{abstract}
\end{titlepage}

\starttext
Since the introduction of D-branes~\cite{pol1}, many calculations have 
appeared~\cite{pol} dealing with one loop amplitudes between various parallel D-branes. Some calculations involved more complicated boundary conditions, such as for moving 0-branes~\cite{bachas}.  In the case of parallel branes the effects of the compactification were discussed in~\cite{lifsch}. Here we show that compactification does not change one loop vacuum amplitude in the case when a ND direction and then generalize this argument for the case of two D-strings at an angle compactified on $T^2.$  
First we consider a simple situation of the two perpendicular non-intersecting D-strings. Assume that first D-string is along $X^1$ and the second one is along $X^2$. The amplitude can be obtained from~\cite{mypap} by setting $\alpha=\frac{\pi}{2}$:
\be
\begin{array}{c}
{\cal A}=2i{\rm T}\int_0^\infty\,\frac{dt}{t}(2\pi t)^{-\frac{1}{2}}
e^{-t\frac{Y^2}{2\pi^2}}\left(\frac{2\pi}{\theta_1'(0|\tau)}
\right)^3\theta_1^{-1}(-\frac{\tau}{2}|\tau)
\times \\ \\ \times \left\{
(-\theta_2^3(0|\tau)\theta_2(-\frac{\tau}{2}|\tau)+\theta_3^3(0|\tau)\theta_3(-\frac{\tau}{2}|\tau)-\theta_4^3(0|\tau)\theta_4(-\frac{\tau}{2}|\tau) 
\right\} . \label{eq:ampl}
\end{array}
\ee
Here $\tau=\frac{it}{2\pi}$. This amplitude depends on the minimal distance between the branes $Y^2=Y^iY^i$, where $i=3...9.$ In obtaining such an amplitude one uses the fact that the fundamental string between these D-string has ND and DN boundary conditions in the directions $X^1$ and $X^2$. The corresponding mode expansion does not contain neither momentum nor winding modes in these two directions. This suggest that the amplitude does not change under the compactification of the direction of these D-strings, i.e. the amplitude between two non-intersecting D-strings compactified on $T^2\times{\cal M}^{7,1}$ with winding numbers $(1,0)$ and $(0,1)$ is the same as the amplitude between the two non-intersecting perpendicular D-strings in ${\cal M}^{9,1}.$
\par
The proof of this conjecture amounts to showing that the summation over the surfaces in this case does not contain topologically different worldsheets from the case of un-compactified D-strings.
\begin{figure} 
\begin{center} \PSbox{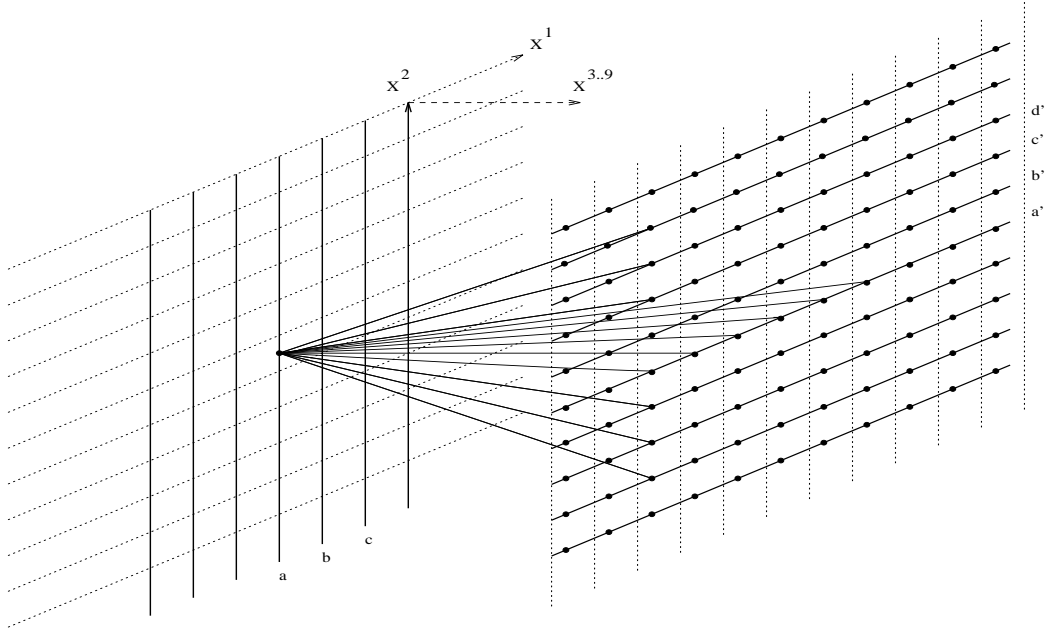 hscale=60 vscale=50}{5in}{3in} \end{center}
\caption{Two nonintersecting D-strings with winding numbers $(0,1)$ and $(1,0)$ on $T^2$ in the covering space of $T^2\times{\cal M}^{7,1}$.}
\end{figure} Figure 1 depicts the case in question in the covering space.
\par
 There are two $(X^1,X^2)$ planes described by the equations 
\be
X^i=0,\; i=3,...,9 \label{eq:lplane}
\ee and 
\be
\begin{array}{l}
X^3=Y\\
X^i=0,\; i=4,...,9. \label{eq:rplane}
\end{array}
\ee
Directions $X^1$ and $X^2$ are compact so that the points
\be
X\sim X+{\cal T}_{mn}
\ee
are identified with
${\cal T}_{mn}=(2\pi n R_1,2\pi m R_2,0,...,0).$
Solid lines the plane~(\ref{eq:lplane}) denoted by $(a,b,c,...)$ in the figure represent identified D-strings described by the equation
\be
X^1=2\pi n R_1 ,\; n\in{\cal Z}
\ee
and the D-strings in the plane~(\ref{eq:rplane}), $(a',b',c',...)$  are given by
\be
X^2=2\pi m R_2 ,\; m\in{\cal Z}
\ee  
 and the dotted lines separate identified strips in the perpendicular direction. 

In order to quantize this system we need to consider the collection of all fundamental strings with one common endpoint
$y^\mu_1$ on a particular D-string in the plane~(\ref{eq:lplane}) and the other endpoints, $y^\mu_{2,mn}=y^\mu_2+{\cal T}_{mn}, \forall m,n\in{\cal Z}$ on the D-strings in the plane~(\ref{eq:rplane}). 

Clearly, a fundamental string with no excitations (thin solid straight line in the figure 1) can have any length greater than the shortest distance between D-strings,
\be
|Y|\leq |y^\mu_1-y^\mu_{2,mn}|<\infty.
\ee
Therefore the entire picture can be replaced by just two non-intersecting perpendicular D-strings . This excludes nontrivial worldsheets that would arise from winding modes, which are not present for ND conditions\thanks{The are present in the amplitude between the two 0-branes, where we have to sum over the discrete set of the possible lengths of the non-excited string connecting 0-branes, which corresponds to the sum over the winding number.} . 

Now we need to show that the compactification along the D-strings, i.e. the identification $X\sim X+{\cal T}_{0n}$ in the plane~(\ref{eq:lplane}) and $X\sim X+{\cal T}_{n0}$ does not change the amplitude. If we move the endpoints of the initially unwound fundamental string by 
\be
\begin{array}{l}
{y'}^\mu_1=y^\mu_1+{\cal T}_{01}\\
{y'}^\mu_2=y^\mu_2+{\cal T}_{10}
\end{array}
\ee
then the string cannot be identified with the original one, since its winding number has changed,
\be
(0,0)\longrightarrow (1,1),
\ee
thus excluding the possibility of nontrivial closed worldsheets associated with momentum modes. In fact in general if we move the endpoints by
\be
\begin{array}{l}
{y'}^\mu_1=y^\mu_1+{\cal T}_{0n}\\
{y'}^\mu_2=y^\mu_2+{\cal T}_{m0}
\end{array}
\ee
along the D-strings, the winding number of the fundamental string changes:
\be
(0,0)\longrightarrow (m,n),
\ee
and the resulting string cannot be identified with the original one for $m^2+n^2\not= 0.$\thanks{However, it seems that a vacuum 
\be
{\rm vacuum}\longrightarrow {\rm closed\; string\; with\; winding\; number}\; 
w=(m,n)
\ee
is an allowed process in this case.}
To illustrate this point better, we consider two parallel interacting D-strings in $S^1\times{\cal R}^9,$ (figure 2) where the direction of the D-strings is compact.
\begin{figure} 
\begin{center} \PSbox{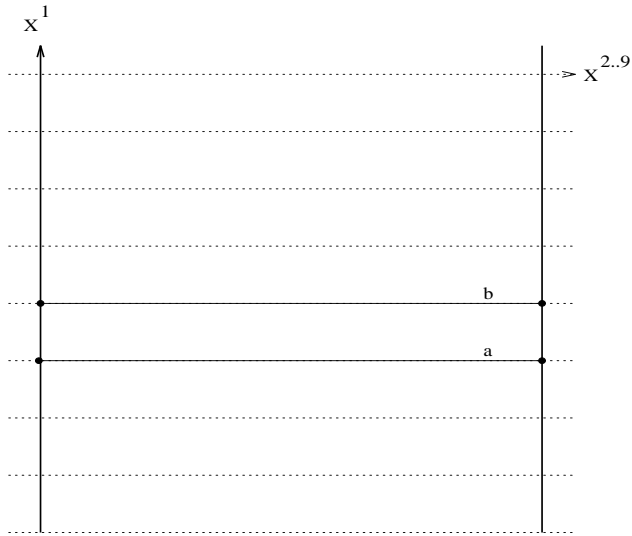 hscale=70 vscale=60}{3in}{2in} \end{center}
\caption{Two parallel D-strings in the covering space of $S^1\times{\cal M}^{8,1}.$}
\end{figure}
 Here, the identification is $X\sim X+{\cal T}_n,$ where ${\cal T}_n=
(2 \pi n R_1,0,...,0).$ Starting form the fundamental string with winding number 0 we move both endpoints by ${\cal T}_1$ and arrive to the same unwound string. (In the figure 2 fundamental strings $a$ and $b$ are identified.) This creates the additional closed worldsheets topologically different from the ones appearing in the case of two parallel D-strings in the non-compact space
 and this is what makes the amplitude different from un-compactified case- the momentum is now discrete.\thanks{The net force is zero, since this is a BPS state.} 
\par
Therefore, we see that the possibility of topologically non-trivial worldsheets is always related to the presence of zero modes for the one loop vacuum amplitude. One should emphasize that this is only the property of the vacuum to vacuum amplitude. In the case of the scattering of the four fundamental strings with one end on a D-string and the other in a D3-brane the amplitude does change if we compactify a ND direction. This is due to the fact that a 3-3 fundamental string is created in the one loop process, and the center of mass of this string can travel around the compact direction in the 3-brane any number of times before the 3-3 string splits into the two 1-3 strings in the final state.~\cite{samir}
\par
Now returning to perpendicular D-strings compactified on $T^2$, if we apply T-duality in the direction of one of the D-strings, we get a system of a 0-brane interacting with a 2-brane, and here again, the amplitude is independent on whether 2-brane is wrapped around $T^2$ or not.  
\par
More generally, if we take a system of two non-parallel non-intersecting D-strings as in~\cite{mypap} at such an angle that when compactified on a $T^2$ the strings will have certain definite winding numbers, the amplitude will be multiplied by the intersection number of the corresponding cycles on $T^2.$ If the first D-string is along $(p,q)$ 1-cycle ${\cal C}_1=p\alpha+q\beta$, where $\alpha,\,\beta$ are homology basis 1-cycles on $T^2$ and the second one is along the $(p',q')$ cycle ${\cal C}_2$, then the intersection number $^{\#}{\cal C}_1\cdot{\cal C}_2$ is given by 
\begin{equation}
^{\#}{\cal C}_1\cdot{\cal C}_2=\left|\begin{array}{cc}
p&p'\\
q&q'
\end{array}\right|,
\end{equation}
since $^{\#}\alpha\cdot\beta=1$ and $^{\#}\beta\cdot\alpha=(-1)^{k(n-k)} \,\left({^{\#}\alpha\cdot\beta}\right)$ for $k$---cycle $\alpha$ and $(n-k)$--cycle $\beta$ on ${\cal M}^n.$ 
The argument for the amplitude given above for two D-strings with $(1,0)$ and $(0,1)$ winding numbers works for each intersection separately, therefore the total answer is given by
\be\label{afin}
{\cal A}_{(p,q),(p',q')}={\cal A}\; \left(^{\#}{\cal C}_1\cdot{\cal C}_2\right),\;\; ^{\#}{\cal C}_1\cdot{\cal C}_2\not= 0,
\ee
where ${\cal A}$ was given in (\ref{eq:ampl}). If $^{\#}{\cal C}_1\cdot{\cal C}_2=0$ then the amplitude involves the standard summation over the winding modes. 
As a concluding remark one may notice that in general if the angle between the D-strings is arbitrary, then after compactification on a $T^2$ they may not form closed cycles, and will fill the entire torus. The intersection number will become infinite and one may expect after taking an apropriate limit in (\ref{afin}) to reproduce an amplitude between two D2-branes compactified on the same $T^2.$
\subsection*{Acknowledgments}

I would like to thank Amer Iqbal, Samir Mathur and Barton Zwiebach for useful discussions.  This work is supported in part by funds provided by the U.S. Department of Energy (D.O.E.) under cooperative research agreement DE-FC02-94ER40818.

\end{document}